\documentclass[reprint, prl, superscriptaddress, amsmath, amssymb, floatfix]{revtex4-2}
\usepackage{graphicx, dcolumn, bm, svg, upgreek, xfrac, float, mathtools, hyperref}

\begin{document}

\title{Engineering the impact of phonon dephasing on the coherence of a WSe$_{2}$ single-photon source via cavity quantum electrodynamics}
\date{\today}

\author{Victor~N.~Mitryakhin}
    \affiliation{Institute of Physics, Carl von Ossietzky University of Oldenburg, 26129 Oldenburg, Germany}
\author{Alexander~Steinhoff}
    \affiliation{Institute for Theoretical Physics and Bremen Center for Computational Material Science, University of Bremen, 28334 Bremen, Germany}
\author{Jens-Christian~Drawer}
    \affiliation{Institute of Physics, Carl von Ossietzky University of Oldenburg, 26129 Oldenburg, Germany}
\author{Hangyong~Shan}
    \affiliation{Institute of Physics, Carl von Ossietzky University of Oldenburg, 26129 Oldenburg, Germany}
\author{Matthias~Florian}
    \affiliation{University of Michigan, Department of Electrical Engineering and Computer Science, Ann Arbor, Michigan 48109, USA}
\author{Lukas~Lackner}
    \affiliation{Institute of Physics, Carl von Ossietzky University of Oldenburg, 26129 Oldenburg, Germany}
\author{Bo~Han}
    \affiliation{Institute of Physics, Carl von Ossietzky University of Oldenburg, 26129 Oldenburg, Germany}
\author{Falk~Eilenberger}
    \affiliation{Friedrich Schiller University Jena, Jena, Germany}
\author{Sefaattin~Tongay}
    \affiliation{Materials Science and Engineering, School for Engineering of Matter, Transport and Energy, Arizona State University, Tempe, 85287, Arizona, USA}
\author{Kenji~Watanabe}
    \affiliation{Research Center for Functional Materials, National Institute for Materials Science, 1-1 Namiki, Tsukuba 305-0044, Japan}
\author{Takashi~Taniguchi}
    \affiliation{International Center for Materials Nanoarchitectonics, National Institute for Materials Science, 1-1 Namiki, Tsukuba 305-0044, Japan}
\author{Carlos~Ant\'on-Solanas}
    \affiliation{Depto. de Física de Materiales, Instituto Nicolás Cabrera, Instituto de Física de la Materia Condensada, Universidad Autónoma de Madrid, 28049 Madrid, Spain}
\author{Ana~Predojevi\'{c}}
    \affiliation{Institute of Physics, Carl von Ossietzky University of Oldenburg, 26129 Oldenburg, Germany}
    \affiliation{Department of Physics, Stockholm University, 10691 Stockholm, Sweden}
\author{Christopher~Gies}
    \affiliation{Institute for Theoretical Physics and Bremen Center for Computational Material Science, University of Bremen, 28334 Bremen, Germany}
\author{Martin~Esmann}
    \affiliation{Institute of Physics, Carl von Ossietzky University of Oldenburg, 26129 Oldenburg, Germany}
\author{Christian~Schneider}
    \affiliation{Institute of Physics, Carl von Ossietzky University of Oldenburg, 26129 Oldenburg, Germany}
    
\begin{abstract}

Emitter dephasing is one of the key issues in the performance of solid-state single photon sources. Among the various sources of dephasing, acoustic phonons play a central role in adding decoherence to the single photon emission. Here, we demonstrate, that it is possible to tune and engineer the coherence of photons emitted from a single WSe$_2$ monolayer quantum dot via selectively coupling it to a spectral cavity resonance. We utilize an open cavity to demonstrate spectral enhancement, leveling, and suppression of the highly asymmetric phonon sideband, finding excellent agreement with a microscopic description of the exciton-phonon dephasing in a truly two-dimensional system. Moreover, the impact of cavity tuning on the dephasing is directly assessed via optical interferometry, which points out the capability to utilize light-matter coupling to steer and design dephasing and coherence of quantum emitters in atomically thin crystals. 

\end{abstract}

\maketitle


In contrast to excitations in isolated atoms, an appropriate  understanding of the light-matter coupling of excitons in solid-state quantum emitters cannot be developed without considering the crucial impact of the lattice environment.  In the context of emission of quantum light, it is the coupling of excitons to acoustic phonons in most semiconductors that manifests itself via the formation of spectral side-bands and replica-like modes \cite{PhysRevB.83.041304, PhysRevLett.101.067402, PhysRevLett.118.253602}. 
 
While such phenomena can be directly employed in quantum-optomechanic experiments \cite{Barzanjeh2022, RevModPhys.86.1391, Chu2018, Kettler2021, WeiSS:21, Pirkkalainen2015, doi:10.1021/nl501413t}, they typically impede the coherence of the emitted light in material-specific manners. Atomically thin crystals belong to the most intriguing class of quantum materials utilized in modern photonic research \cite{Liu:19}. Due to their ultimate thinness, the correlations of carriers are strongly enhanced and dominate their optical response. In the presence of random or engineered local strain, the formation of luminescent hot spots has been verified \cite{he_single_2015, Branny2017}, displaying non-classical light emission. Such quantum dots (QDs) can be found in the majority of thin-layer transition metal dichalcogenide (TMDC) crystals. However, they are particularly pronounced in WSe$_2$ mono- and bilayers, which have been extensively studied in this regard \cite{koperski_single_2015, he_single_2015, chakraborty_voltage-controlled_2015, srivastava_optically_2015, tonndorf_single-photon_2015}. QDs in TMDCs have recently shifted into the spotlight of quantum photonic research. In contrast to conventional quantum emitters in bulk material, two-dimensional QDs are highly tunable without the need for demanding semiconductor processing \cite{iff_strain-tunable_2019, chakraborty_voltage-controlled_2015}. They can be conveniently integrated into photonic architectures without the restrictions of lattice matching \cite{849b8cabbde64f69a000744dd26ce668}, which can be realized via simple low-cost pick-and-place methods \cite{castellanos-gomez_deterministic_2014}. 

Despite this ease of integrating TMDC QDs into photonic devices, the enhancement of spontaneous emission from monolayer QDs has been demonstrated in a limited number of reports. The Purcell regime has been verified using TMDC-Bragg gratings \cite{iff_purcell-enhanced_2021}, plasmonic structures \cite{tripathi_spontaneous_2018, cai_radiative_2018, luo_deterministic_2018}, and most recently in tunable open-optical cavities \cite{doi:10.1021/acs.nanolett.3c02584}. 
 
While it is clear that dephasing channels compromise and dictate the coherence of the emitted photons, they are yet to be exhaustively studied.  While it is possible to engineer crystal superlattices on the nano-scale to suppress coupling to phonons, such approaches are typically technologically too demanding to be routinely implemented \cite{PhysRevLett.43.2012, PhysRevLett.89.227402, PhysRevB.33.1516, PhysRevB.33.2897, Jusserand1989, PhysRevB.64.033306}. Up to now, spectral wandering has been identified as one of the main sources of decoherence in the emission of 2D crystal QDs \cite{doi:10.1021/acs.jpclett.9b02863, White:21,  brotons-gisbert_coulomb_2019}. In addition, it was shown that due to the large Huang-Rhys factor of WSe$_2$, the coupling to phonons is extraordinarily strong and induces a very rich phonon-sideband spectrum \cite{PhysRevB.100.125308}. 

    \begin{figure*}[t]
            \includegraphics{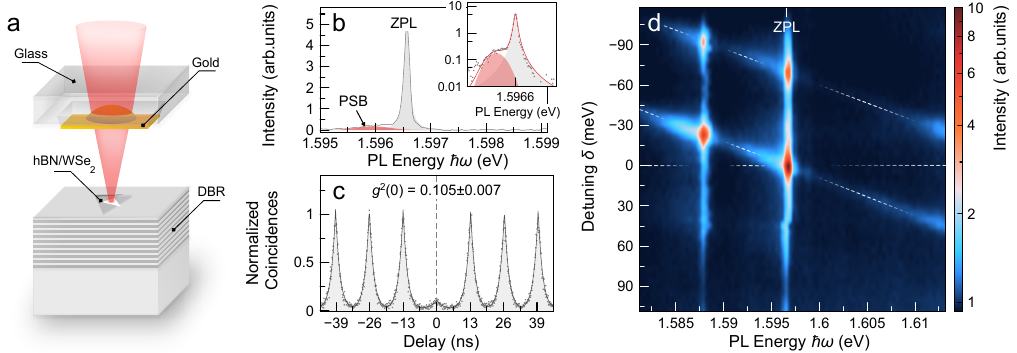}
            \caption{\label{fig:one} (a) Schematics of an open cavity device under optical excitation. A beam is focused through the top concave mirror onto the monolayer and exciting the QD thus triggering the emission of single photons. The generated single photons escape the cavity through the top mirror. (b) Low-resolution PL spectrum of the QD at 3.2 K recorded without the cavity top mirror.  (c) Second-order correlation function. The separation of the peaks is inverse proportional to the laser repetition rate. The peaks have been fitted with an ensemble of double exponential functions $A\cdot e^{\sfrac{-|x-x_0|}{t}}$. (d) PL spectra as a function of relative detuning of a cavity resonance from the emission energy at appr. 1.596 eV (labeled ZPL). The intensity (color coded) is plotted in log$_{10}$ scale. The inclined dashed white lines indicate the cavity modes. As a cavity mode sweeps through emission energies, the emission is gradually enhanced making both the ZPL and PSB (the tail to the left of the ZPL) more pronounced.}
    \end{figure*}

 Here, we demonstrate that the unique shape of the phonon sideband in WSe$_2$ QDs can be exploited to engineer its coherence properties via emitter-cavity coupling. Specifically, by tuning the optical resonance of our cavity system to match either the zero-phonon emission line (ZPL) or the phonon sideband (PSB), we control the individual contributions of these two features to the overall emission spectrum. As a result, the temporal coherence of the single photon emission can be manipulated, changing from a slowly decaying single exponential trace to a biexponential-like trace as the cavity resonance is tuned across the emission spectrum. 


The WSe$_2$ monolayer is placed on the surface of a mirror that is part of an asymmetric plano-concave open cavity (depicted in Fig. \hyperref[fig:one]{\ref*{fig:one}(a)}). The cavity consists of two mirrors that are freely movable. The monolayer hosts QDs arising from crystalline defects in the monolayer. The emission from one of the QDs is investigated. The QD is driven by a Ti:Sapphire laser operating at 1.722 eV. The further details of the cavity system can be accessed in Supplementary Information (SI).

Prior to the studies of photoluminescence (PL) from the QD inside the cavity, the QD emission without a cavity is investigated (see Fig. \hyperref[fig:one]{\ref*{fig:one}(b)}). A PL spectrum of the  QD features a pronounced ZPL and a lower energy PSB. Fig. \hyperref[fig:one]{\ref*{fig:one}(b)} shows the ZPL spectrum located at appr. 1.596 eV, with a linewidth of 110$\pm$3 $\upmu$eV (spectrometer resolution-limited). The linewidth of the PSB is  0.7$\pm$0.1 meV. The ZPL and PSB are separated by 0.6 meV resulting in an asymmetric shape of the emission spectrum (see the inset of Fig. \hyperref[fig:one]{\ref*{fig:one}(b)}, plotted in logarithmic scale). 

To verify the quantum character of the PL emission, we implement a second-order correlation measurement (see Fig. \hyperref[fig:one]{\ref*{fig:one}(c)}), resulting in $g^2(0)=$ 0.105$\pm$0.007. The measurement is carried out in a resonant cavity-QD configuration, profiting from the significant signal enhancement.  

\begin{figure}[t]
        \includegraphics{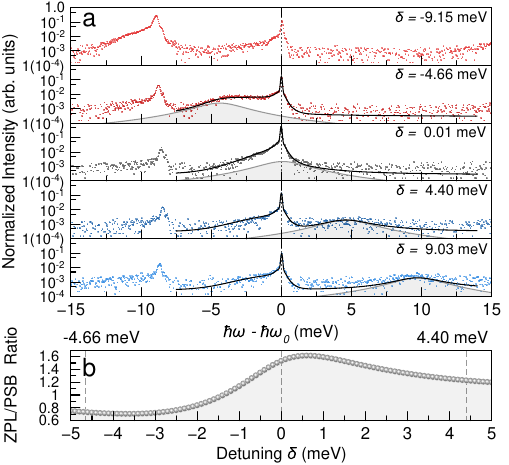}
        \caption{\label{fig:two} (a) PL spectra (y-axis is in log$_{10}$ scale) for selected cavity-emitter detunings [-9.15, -4.66, 0.01, 4.40, 9.03] meV excited by the Ti:Sapphire laser in continuous wave regime at 720 nm wavelength. The position of the cavity is indicated by the grey shaded region. The solid black lines arise from the theoretical model. From the fit we obtain a lattice temperature of $4.6$ K. (b) Ratio between the intensities of the ZPL and the PSB emission  as a function of cavity-ZPL detuning.}
    \end{figure}

The resonant enhancement of photon flux is a direct consequence of the modification of the emitter decay rate and light distribution in the cavity via the Purcell effect \cite{purcell_spontaneous_1995}. The modification of the luminescence while sweeping the cavity resonance through the QD ZPL is shown in Fig. \hyperref[fig:one]{\ref*{fig:one}(d)}. These PL spectra are plotted vs the cavity-emitter detuning $\delta = \hbar\omega_{cav} - \hbar\omega_0$, where $\hbar\omega_{cav}$ is the energy of the cavity resonance and $\hbar\omega_0$ is the ZPL energy (at 1.596 eV, see Fig. \hyperref[fig:one]{\ref*{fig:one}(d)}, and extracted PL spectra for several detunings in Fig. \hyperref[fig:two]{\ref*{fig:two}(a)}). When the ZPL is in resonance with the cavity, the intensity of the emission is enhanced by more than a factor of 6 compared to the off-resonant case. The PSB is affected by the cavity tuning: when the cavity is positively detuned, the contribution of the PSB to the PL spectrum is almost negligible and the ZPL dominates. For negative detunings, the PSB contribution is notable and manifests itself as a distinct spectral feature in  Fig. \hyperref[fig:one]{\ref*{fig:one}(d)}. We study this phenomenon to quantify the interplay between the ZPL and PSB integral intensities for cavity-emission tuning by applying a fitting procedure based on the independent boson model for quantum emitters interacting with phonons in the two-dimensional monolayer \cite{PhysRev.139.A1965}. We include coupling of the local emitter to two-dimensional longitudinal acoustic phonons \cite{klein_site-selectively_2019} and a localized phonon mode \cite{Wigger_2019}. The optical susceptibility of the QD-phonon system is 
    \begin{equation}
        \begin{split}
            \chi(t)=i\theta(t)e^{-i\omega_{0}t+\Phi(t)-\Gamma_{\textrm{inhom}}^2 t^2}
        \end{split}
    \label{eq:susceptibility}
    \end{equation}
\noindent with $\Gamma_{\textrm{inhom}}$ being the inhomogeneous broadening and $\Phi(t)$ being the phonon dephasing integral:
    \begin{equation}
        \begin{split}
            \Phi(t) = \sum_j\int_0^{\infty}d\omega \frac{J_j(\omega)}{\pi\omega^2}
            \Big\{&\textrm{coth}\Big(\frac{\hbar\omega}{2 k_{\textrm{B}}T}\Big)\big[\textrm{cos}(\omega t)-1\big]\\
            -i&\textrm{sin}(\omega t)\Big\}\,.
        \end{split}
    \label{eq:Phi_t}
    \end{equation}
\noindent Here, $J_j(\omega)$ is the spectral density of phonon branch $j$ as detailed in the SI.
The corresponding emission spectrum is obtained by inverting the absorption spectrum at the ZPL:

    \begin{equation}
        \begin{split}
            I_{\textrm{QD-PSB}}(\omega) =\ &
            \alpha(2\omega_0-\omega)\\ =\ &
            \textrm{Im}\,\Big\{
            \int_{-\infty}^{\infty}dt\,\chi(t)e^{i(2\omega_0-\omega)t}
            \Big\}\,.
    \end{split}
    \label{eq:em_spec}
    \end{equation}

The cavity is explicitly accounted for by multiplying the QD-sideband emission spectrum with a lineshape function $L(\omega)$ describing the photonic density-of-states as discussed in the SI. Translating the modeled emission spectra to experimentally observed intensities requires an additional scaling factor $a$:
    \begin{equation}
        \begin{split}
            I(\omega)=
            a \,\big(I_{\textrm{QD-PSB}}(\omega) + I_{\textrm{BG}}\big)\, L(\omega)     \,.      
        \end{split}
    \label{eq:em_spec_final}
    \end{equation}
Here, background $I_{\textrm{BG}}$ is added accounting for a possible contribution of a low-energy tail of another emitter and emission of the monolayer free exciton.

We simultaneously fit the data sets for several detunings with the same emitter-phonon coupling parameters.
To account for experimentally determined fluctuations, only small variations of the ZPL position, the cavity parameters and the scaling factor $a$ are allowed with detuning. The obtained fit parameters are collected in the SI. The resulting spectral fittings are shown in Fig. \hyperref[fig:two]{\ref*{fig:two}(a)}. For the lowest detuning (top panel), the cavity spectrally overlaps with the second QD (located at 1.588 eV), which hinders fitting the spectrum.  
    \begin{figure*}[t]
        \includegraphics{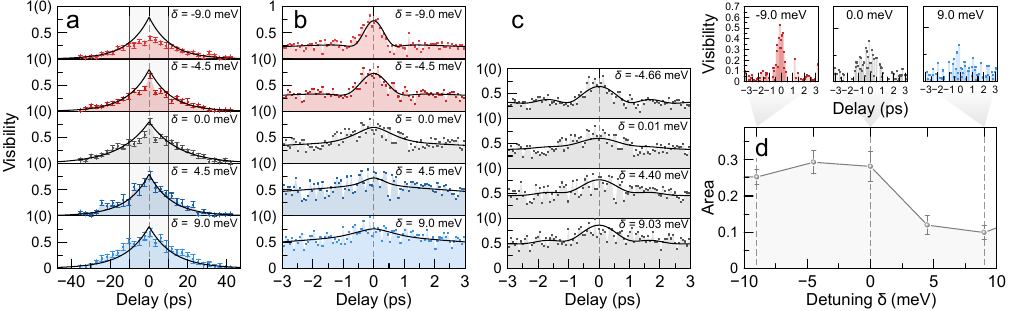}
        \caption{\label{fig:three} Interference visibility: (a) for a large range and (b) for a small range of the time delay between the arms of the interferometer. Markers indicate the data obtained in the experiment, solid lines indicate fit functions. In panel  (a), the fit is based on the tails outside of the shaded area, see explanation in the text. (c) Solid lines represent the visibility data calculated as modulus of the Fourier transformed fit functions from Fig. \hyperref[fig:two]{\ref*{fig:two}(a)} including a bandpass filter model. The data is normalized to match the observed visibility of the experiment. (d) Area of the visibility (from which the slow decaying trace is subtracted) as function of the cavity-emitter detuning. For the points corresponding to the detunings -9, 0 and 9 meV, the leftover fragment is plotted in panels on top and represents the contribution of the PSB to the overall interference visibility.}
    \end{figure*}
As detailed in the SI, our model can be analytically solved yielding a decomposition of the spectrum into phonon-assisted processes of arbitrary order. This allows us to quantify the ZPL/PSB contribution to the emission. Due to the two-dimensional nature of acoustic phonons, the carrier-phonon coupling efficiency does not vanish at small momenta as it is the case in  three dimensions \cite{krummheuer_theory_2002}. Hence the ZPL spectrally overlaps with low-energy higher-order processes that involve the absorption and emission of phonons. We sum up the zero-phonon contributions and the higher-order contributions with energies smaller than the full-width-at-half-maximum of the inhomogeneous emitter broadening to obtain an effective ZPL.  A weighting of ZPL and PSB with the cavity lineshape at different detunings quantifies the ZPL/PSB interplay ( see Fig. \hyperref[fig:two]{\ref*{fig:two}(b)}): the resulting extracted ratio of the ZPL to the PSB is approximately 1.6 for the positively detuned cavity, indicating a regime that is dominated by the ZPL. For the negatively detuned case, it approaches 0.7,  suggesting a tremendous impact of the PSB on the coherence of single photons emitted from the QD-cavity device under such detuning conditions. 

To account for the limitations of conventional spectroscopy and to directly verify the impact of these phenomena on photon coherence, we extend our study with interferometric measurements in the time domain. Here, we filtered the QD-cavity emission using a bandpass filter with an approximate bandwidth of 2.5 meV (which only passes the QD emission) and conducted a set of first-order correlation  measurements using a Michelson interferometer as we change the emitter-cavity detuning (Fig. \hyperref[fig:three]{\ref*{fig:three}(a-b)}). Scanning the relative phase between the arms of the interferometer, we obtain interferograms displaying intensity oscillations. Yielding the value of the first-order correlation function for a given temporal delay in the interferometer, the interference visibility $v$ of these oscillations is extracted as   $v = (I_{max}-I_{min})/(I_{max}+I_{min})$, where $I_{max/min}$ is the upper/lower envelope of the interferogram signal.

In Fig. \hyperref[fig:three]{\ref*{fig:three}(a)}, we plot the resulting extracted interference visibility as we coarsely scan our interferometer. The measurements reveal a significant influence of the emitter-cavity detuning on the first order coherence of the system. While all visibility traces decay with constant coherence time at longer time-scales, at very short time scales, the interference visibilities for negative detunings are reduced and deviate from the simple exponential decay. This phenomenon is further investigated in Fig. \hyperref[fig:three]{\ref*{fig:three}(b)} with finer time resolution while showing the visibility in the vicinity of the zero path difference between the two interferometer arms. 

The visibility plots exhibit a pronounced fast decay due to the PSB (which approximately is of Gaussian shape, yielding the correct quadratic behavior at small times) and the slowly decaying background that we attribute to the ZPL emission \cite{krummheuer_theory_2002, zimmermann_dephasing_2002}. It is in agreement with a recent report on the coherence of a similar WSe$_2$ based QDs \cite{von_Helversen_2023}. Crucially, the impact of the pronounced fast decay clearly depends on the QD-cavity detuning, whereas the slow decay remains widely unaffected. 

Since the interferograms display the temporal coherence profile of the emission, they are connected to the spectral shape of the emission via Fourier transformation (FT). Hence, we can harness the power of our fully microscopic model for the QD-phonon system and express the theoretically obtained visibilities of photon coherence as: 

    \begin{equation}
        \begin{split}     
        v(t)=N\Big|\int_{-\infty}^{\infty}d\omega\,I(\omega)H(\omega)e^{i\omega t}\Big|\,.
        \end{split}
    \label{eq:em_spec_FT}
    \end{equation}

\noindent For the emission spectrum $I(\omega)$ we directly use the fit functions shown in Fig. \hyperref[fig:two]{\ref*{fig:two}(a)}. The function $H(\omega)$ models a bandpass filter and $N$ is a normalization constant, for details see the SI. To adjust $H(\omega)$ and $N$, we simultaneously fit the experimental visibility for several detunings (except for the -9.15 meV, for the same reason as for Fig. \hyperref[fig:two]{\ref*{fig:two}(a)}) with the same bandpass parameters. We assume that the deviation of detunings used for interferometry and for spectroscopy are negligible.  

Numerical fits to the experimental visibility data using Eq.~(\ref{eq:em_spec_FT}) are shown in Fig. \hyperref[fig:three]{\ref*{fig:three}(c)}, where the cavity detuning is varied according to the theoretical spectra shown in Fig. \hyperref[fig:two]{\ref*{fig:two}(a)}. We find an overall good agreement with the experimentally obtained temporal coherence, which corroborates the consistency of our microscopic model. The most notable feature that cannot be extracted directly from the experimental data is the pronounced oscillation of the signal at negative detuning with a period of about $1$ ps. We attribute it to an interference between the emitter and the cavity-enhanced PSB.

To extend our quantitative analysis and extract coherence times, we fit the experimental data with a more phenomenological model based on the FT of a typical emission spectrum of such QDs. We approximate the intensity of the signal recorded in the Michelson interferometer as
    \begin{equation}\label{eq:interferogram}
        I(t) = I_0 + \frac{A_1}{\tau_1} e^{-\frac{|t|}{\tau_1}}\cos{(\omega_1t)} +\frac{A_2}{\tau_2} e^{-\frac{t^2}{\tau_2^2}}\cos{(\omega_2t)}.
    \end{equation}
\noindent where $t$ is the delay between the two arms of the interferometer, $I_0$ is the intensity of the signal at the input of the interferometer, and $A_{1,2}$, $\tau_{1,2}$, $\omega_{1,2}$  are the amplitudes, decay times and the frequencies  of the individual components, respectively. The extraction of the envelopes is possible by considering the interference of two plane waves, resulting in
    \begin{equation}\label{eq:interferogram-envelope}
        \begin{split}
         I_{max,min}(t) = I_0\pm\biggl\{\frac{A_1^2}{\tau_1^2} e^{-2\frac{|t|}{\tau_1}} + \frac{A_2^2}{\tau_2^2} e^{-2\frac{t^2}{\tau_2^2}} + \\
         + 2A_1A_2 e^{-\frac{|t|}{\tau_1}}e^{-\frac{t^2}{\tau_2^2}}\cdot \cos{(|\omega_1-\omega_2|t)}\biggl\}^{\sfrac{1}{2}}.
        \end{split}
    \end{equation}
\noindent Since the upper and the lower envelopes are symmetric according to Eq. \hyperref[eq:interferogram-envelope]{(\ref*{eq:interferogram-envelope})}, the visibility trace is expressed  as $v\propto |I_{max}|$. 

In Fig. \hyperref[fig:three]{\ref*{fig:three}(a)}, we perform fitting of the visibility using the function proposed in Eq. \hyperref[eq:interferogram-envelope]{(\ref*{eq:interferogram-envelope})} without the Gaussian decay term and excluding the range of [-10, 10] ps (shown as shaded area), since the double exponential term is dominant at longer time scales.  Our fitting procedure reveals a coherence time associated with the ZPL of $11.8\pm0.4$ ps, which is consistent with the spectral ZPL linewidth of $110$ µeV. The value of the coherence time is two orders of magnitude smaller than the characteristic radiative lifetime, see the data presented in the SI.

In Fig. \hyperref[fig:three]{\ref*{fig:three}(b)}, the data is fitted with a function proportional to Eq. \hyperref[eq:interferogram-envelope]{(\ref*{eq:interferogram-envelope})} with $\tau_1$ fixed to 11.8 ps. From the fits for negative detunings, we extract an average fast component $\tau_2=0.85\pm0.15$ ps, which we adopt as fixed value for the zero and positive detuning fits. 

The described procedure allows to quantify the PSB contribution to the visibility decay as a function of cavity-emitter detuning. This contribution is most straightforwardly reflected by its relative area below the visibility traces. We extract this area by subtracting the slow decaying (ZPL) trace to obtain the leftover fragments in the top panel of Fig. \hyperref[fig:three]{\ref*{fig:three}(d)}. As shown in the bottom panel, the area suggests that the phonon impact on the first order temporal coherence is of great significance for negative as well as zero detuning conditions, but can be considerably suppressed for positive detunings.


In summary, we provide a novel pathway to  cavity-control the optical properties of quantum emitters in general, and WSe$_2$ QDs in particular. While previous works specifically utilized the coupling of QDs to optical cavities to enhance the photon flux, as well as to engineer the spontaneous emission lifetime, here, we verify that the cavity-emitter detuning directly influences the emitter coherence time (T$_{2}$). This significant effect is prominently exposed in the spectral as well in the temporal domain. The way to gain control of the emitter coherence utilizes selective enhancement of the emission of the zero phonon line versus the rapidly dephasing phonon sideband (which is unavoidable and material-specific). Indeed, this method does not require high-Q cavities and works particularly well in the case of WSe$_2$ QDs featuring a very asymmetric PSB. Our QDs still suffer from a rapidly dephased ZPL that sets limits to the overall temporal coherence. We attribute this effect to the effective dephasing provided by the emitter-phonon interaction in a truly two-dimensional system. Our methodology to engineer the impact of the phonon collisions on the coherence of the emitted light beam is universal and highly relevant in the context of realizing quantum emitters in 2D systems. It will be possible to apply it to the next generation WSe$_2$ QDs with reduced dephasing, possibly based on optimized crystals with advanced charge control as well as excited using resonant driving schemes, which will pave the way towards coherent TMDC single photon sources. 

\begin{acknowledgments}

The authors acknowledge funding from the German federal ministry of education and research (BMBF) within the projects ‘TUBLAN’ and ‘EQUAISE’. The project ‘EQUAISE’ is funded within the QuantERA program. S.T. acknowledges primary support from NSF DMR 2111812 for materials development, NSF GOALI 2129412 for scaling, and NSF ECCS 2111812 fabrication. C.G. acknowledges funding from the Deutsche Forschungsgemeinschaft (DFG) priority program SPP2244. We acknowledge partial support for DOE-SC0020653 (materials texture development), NSF ECCS 2052527 for electronic and NSF DMR 2206987 for magnetic purity tests. C.A.S. acknowledges the support from the Comunidad de Madrid fund “Atraccion de Talento, Mod. 1”, Ref. 2020‐T1/IND‐19785 and the project from the Ministerio de Ciencia e Innovación PID2020113445GB-I00. M.F. acknowledges support by the Alexander von Humboldt foundation. A. P. acknowledges the Helene Lange Visiting Professorship program. M.E. acknowledges funding from the University of Oldenburg through a Carl von Ossietzky Young Researchers' Fellowship.  Financial support by the Niedersächsisches Ministerium für Wissenschaft and Kultur ("DyNano") is acknowledged. C. S. acknowledges the support by the German Research Foundation (INST184/220-1 FUGG).

\end{acknowledgments}

\bibliography{main}
\end{document}